# Imaging Moiré Excited States with Photocurrent Tunneling Microscopy

*Hongyuan Li*[1, 2, 3, 10], *Ziyu Xiang*[1, 2, 3, 10], *Mit H. Naik*[1, 3, 10], *Woochang Kim*[1, 3], *Zhenglu Li*[1, 3], *Renee Sailus*[4], *Rounak Banerjee*[4], *Takashi Taniguchi*[5], *Kenji Watanabe*[6], *Sefaattin Tongay*[4], *Alex Zettl*[1, 3, 7], *Felipe H. da Jornada*[8, 9], *Steven G. Louie*[1, 3]*, *Michael F. Crommie*[1, 3, 7]* and *Feng Wang*[1, 3, 7]*

[1]Department of Physics, University of California at Berkeley, Berkeley, CA, USA.

[2]Graduate Group in Applied Science and Technology, University of California at Berkeley, Berkeley, CA, USA.

[3]Materials Sciences Division, Lawrence Berkeley National Laboratory, Berkeley, CA, USA.

[4]School for Engineering of Matter, Transport and Energy, Arizona State University, Tempe, AZ, USA.

[5]International Center for Materials Nanoarchitectonics, National Institute for Materials Science, Tsukuba, Japan

[6]Research Center for Functional Materials, National Institute for Materials Science, Tsukuba, Japan

[7]Kavli Energy Nano Sciences Institute at the University of California Berkeley and the Lawrence Berkeley National Laboratory, Berkeley, CA, USA.

[8]Department of Materials Science and Engineering, Stanford University, Palo Alto, CA, USA.

[9]Stanford Institute for Materials and Energy Sciences, SLAC National Accelerator Laboratory, Menlo Park, CA, USA.

[10]These authors contributed equally: Hongyuan Li, Ziyu Xiang, and Mit H. Naik

**Abstract:**

Moiré superlattices provide a highly tunable and versatile platform to explore novel quantum phases and exotic excited states ranging from correlated insulators[1-17] to moiré excitons[7-10,18]. Scanning tunneling microscopy has played a key role in probing microscopic behaviors of the moiré correlated ground states at the atomic scale[1,11-15,19]. Atomic-resolution imaging of quantum excited state in moiré heterostructures, however, has been an outstanding experimental challenge. Here we develop a novel photocurrent tunneling microscopy by combining laser excitation and scanning tunneling spectroscopy (laser-STM) to directly visualize the electron and hole distribution within the photoexcited moiré exciton in a twisted bilayer $WS_2$ (t-$WS_2$). We observe that the tunneling photocurrent alternates between positive and negative polarities at different locations within a single moiré unit cell. This alternating photocurrent originates from the exotic in-plane charge-transfer (ICT) moiré exciton in the t-$WS_2$ that emerges from the competition between the electron-hole Coulomb interaction and the moiré potential landscape. Our photocurrent maps are in excellent agreement with our $GW$-BSE calculations for excitonic states in t-$WS_2$. The photocurrent tunneling microscopy creates new opportunities for exploring photoexcited non-equilibrium moiré phenomena at the atomic scale.

Two-dimensional (2D) moiré superlattices exhibit novel correlated ground and excited states ranging from correlated insulators[1-16] and superconductivity[20-22] to moiré excitons[7-10,18]. Although measurements such as electrical transport[16,22-24] and optical spectroscopy[2,3,25] has revealed mesoscopic properties of the moiré correlated phenomena, knowledge about their microscopic nature is still limited. The microscopic study is particularly important to understand how these correlated phenomena are affected by the interplay between charge, orbitals, and lattices. Scanning tunneling microscopy (STM) has played an indispensable role in exploring the moiré physics at the nanometer scale, which provides real-space information of various correlated grounds states such as Mott insulator[11-15], electron crystal[1,17], and superconductivity[20,21]. However, so far little progress has been made on exploring the moiré excited states at the nanometer scale. Probing individual electrons and holes in the transient excited states is intrinsically challenging since it requires high spatial resolution, charge sensitivity and effective excitation, which is beyond the capability of typical optical spectroscopy techniques. There has been recent effort on using photoemission spectroscopy[11] and electron excitation[12] to microscopically probe the moiré excited states, which, however, still cannot spatially resolve the internal structures of the photoexcited holes and electrons. Here we combine laser excitation with STM (termed laser-STM) to probe the individual electron and hole properties in moiré excited states at the nanoscale. We developed a photocurrent tunneling microscopy (PTM) technique based on the laser-STM, which enabled us to measure the internal microscopic structure of photoexcited states in near-60-degree twisted $WS_2$ (t-$WS_2$) bilayers and reveal the emergence of a new type of charge-transfer moiré excitons. Related laser-STM methods have been used to probe photoexcitation in molecule systems[26-29] and conventional bulk semiconductors[30-35].

Fig. 1a shows an illustration of our gate-tunable t-WS$_2$ device and laser-STM setup (see also Fig. S1). The t-WS$_2$ sits on top of a 49 nm thick layer of hexagonal boron nitride (hBN) which is placed above a graphite back gate. A back gate voltage ($V_{BG}$) is applied between the t-WS$_2$ and the graphite back gate to control the carrier density in the t-WS$_2$. We use a graphene nanoribbon array electrode on the t-WS$_2$ surface to make the sample conductive enough for STM measurements[36]. A bias voltage ($V_{bias}$) is applied to the t-WS$_2$ relative to the STM tip to induce a tunnel current. Details of the device fabrication are included in the Methods section.

Fig. 1b shows a typical STM topography image of the t-WS$_2$ surface and features a moiré superlattice with a moiré period of ~9 nm. Three high-symmetry stacking regions can be seen within the moiré unit cell: the dark (bright) areas correspond to AB ($B^{S/S}$) stacking regions while the intermediate height area is a $B^{W/W}$ stacking region. Model structures for the AB, $B^{S/S}$, and $B^{W/W}$ stacking regions are illustrated in Fig. 1c. This topography agrees well with our calculated height distribution map (Fig. 1d) and arises due to an out-of-plane reconstruction that leads to strong variation in the local interlayer spacing (see SI for details). The $B^{S/S}$ stacking regions have the largest interlayer spacing due to steric hindrance arising from the S atoms of the top and bottom layers facing each other[37]. In-plane reconstruction of the superlattice leads to a noticeably larger area for the low-energy AB stacking region and results in strain redistribution in the two WS$_2$ layers (see SI for details). The final relaxed structure reflects a trade-off between the energy gain from forming large-area AB stacking regions and the energy cost of strain redistribution.

We characterized the single-particle electronic structure of t-WS$_2$ by measuring scanning tunneling spectroscopy and comparing it to theoretical calculations. Fig. 2a shows the calculated electronic band structure of a t-WS$_2$ moiré superlattice with 9nm period including spin-orbit coupling. Here we are mainly concerned with the moiré flat bands at the valence band top (VBT) (labeled v1) and the conduction band bottom (CBB) (labeled c1), which host the lowest-energy holes and electrons, respectively. The computed flat bands arise from a deep triangular quantum well potential due to the inhomogeneous layer hybridization and structural reconstruction[37,38] which yield narrow bandwidths of 1 meV and 5 meV for the v1 and c1 bands, respectively. The v1 band is derived from states near the Γ point in the pristine unit-cell Brillouin zone (BZ) and is two-fold degenerate, while the c1 states are folded from around the Q point of the pristine unit-cell BZ and comprise two closely spaced bands (each with six-fold degeneracy). The small splitting in energy of these two sets of c1 bands is induced by interlayer hybridization. The spatial distribution of these moiré flat bands is experimentally reflected by the dI/dV spectra measured at the $B^{W/W}$ and AB sites (Fig. 2b and 2e). On the valence band side (Fig. 2b) the AB site shows a sharp peak around $V_{bias}$ = -1.75V while the $B^{W/W}$ site exhibits almost no signal until $V_{bias}$ = -1.85V, indicating that the two-fold degenerate v1 band is mainly localized on the AB site. This spatial distribution was confirmed by dI/dV mapping measured at $V_{bias}$ = -1.75V (Fig. 2c) which matches the calculated v1 band's charge density ($\rho$) distribution (Fig. 2d), both of which show maximum density at the AB site. On the conduction band side (Fig. 2e), the lowest energy dI/dV peak appears at the $B^{W/W}$ site at $V_{bias}$ = 0.60V while the AB site exhibits a small signal until $V_{bias}$ = 0.75V, indicating that the c1 band states are mainly localized on the $B^{W/W}$ site. This was also confirmed by dI/dV mapping measured at $V_{bias}$ = 0.62V (Fig. 2f) which matches the calculated c1 bands' charge density distribution (Fig. 2g), both of which show maximum density at the $B^{W/W}$ site. These results show that a strong moiré potential exists in t-WS$_2$ and that the lowest-energy single-particle electron and hole states are spatially separated (more details on the calculations are included in the SI). We note that a moiré superlattice modifies photoexcited states in different ways depending on the strength and shape of the moiré potential. Weak moiré potentials, for example, generate Umklapp scattering of the pristine Wannier excitons that leads to some spatial modulation while retaining their overall Wannier characteristics[39]. Strong moiré potentials, such as those created by strain-induced reconstructions in TMD bilayers, not only induce Umklapp scattering of excitons but can also alter the internal structure of moiré excitons[10].

To experimentally probe the excitonic states we illuminated our t-WS$_2$ device with a 520nm continuous-wave laser focused on the tip-sample tunnel junction (Fig. S1). Real-time feedback control of the laser alignment was performed to maintain a stable laser spot relative to the tip position (see Methods). Photoexcited electrons and holes in t-WS$_2$ relax quickly (on the time scale of ps) to the lowest-energy moiré exciton state through phonon emission[40-42]. Here we set the back gate voltage $V_{BG}$ near 0 to keep the t-WS$_2$ undoped so as to avoid free carrier scattering that could decrease the exciton lifetime[43]. The spatial charge distribution of the lowest-energy (long-lived) moiré exciton state was probed through STM tunneling photocurrent measurement (electron tunneling processes occur on the time scale of ns). Fig. 3a and 3b show the absolute value of the tunnel current (I) as a function of $V_{bias}$ measured at B$^{W/W}$ and AB stacking sites with (blue) and without (orange) laser illumination (the laser power for this measurement was P=600μW and the laser illumination area is around 80 $\mu m^2$). When the laser is off the tunnel current for both stacking sites shows a large semiconducting bandgap for -2V < $V_{bias}$ < 1V. However, when the laser is turned on a photocurrent emerges even when $V_{bias}$ lies in this gap region and the photocurrent response at the B$^{W/W}$ and AB sites show very different behavior.

To investigate the spatial dependence of the tunneling photocurrent, we performed 2D photocurrent mapping with $V_{bias}$ = -0.6V and $V_{BG}$ = 0 as shown in Fig. 3c (laser power was held at P = 600uW). Surprisingly, the photocurrent changes sign at different locations even for fixed $V_{bias}$: positive photocurrent (red) appears at AB sites while negative photocurrent (blue) appears at $B^{W/W}$ sites. This spatially alternating photocurrent polarity provides direct experimental evidence for the emergence of in-plane charge transfer (ICT) moiré excitons.

To better interpret the photocurrent spatial distribution, we performed calculations of the excitonic states of t-WS$_2$ using the *ab initio GW*-BSE[44,45]. Here the exciton wavefunction $\chi_S$ is expressed as a linear combination of single-particle conduction ($c\mathbf{k}_m$) and valence states ($v\mathbf{k}_m$) in the moiré BZ: $\chi_S(\mathbf{r_e}, \mathbf{r_h}) = \sum_{cv\mathbf{k}_m} A^S_{cv\mathbf{k}_m} \psi^*_{v\mathbf{k}_m}(\mathbf{r_h}) \psi_{c\mathbf{k}_m}(\mathbf{r_e})$ where S is the exciton principal quantum number, $\mathbf{k_m}$ is the electron wave vector in the moiré BZ, $\mathbf{r_e}$ and $\mathbf{r_h}$ are the electron and hole coordinates, respectively, $v$ and $c$ label a valence and conduction band, respectively, and $A^S_{cv\mathbf{k}_m}$ are exciton electron-hole expansion coefficients. The exciton states can be calculated (including electron and hole degree of freedom) by state-of-the-art full-spinor *GW*-BSE calculations. However, due to the large number of atoms (~4000) in the t-WS$_2$ moiré unit cell, this calculation is computationally intractable.

To overcome this bottleneck, we developed a new computational algorithm for calculating the electron-hole interaction kernel matrix elements. This involves extending the pristine unit-cell matrix projection (PUMP) method (developed previously for reconstructed monolayers[8]) to bilayer moiré superlattices. In the new scheme, the wavefunction of each valence ($|\psi_{v\mathbf{k}_m}\rangle$) and conduction ($|\psi_{c\mathbf{k}_m}\rangle$) state in the moiré BZ is expressed as a linear combination of many states of the pristine individual layers: $|\psi_{v\mathbf{k}_m}\rangle = \sum_i a_i^{v\mathbf{k}_m} |\Phi^{val}_{a,i\mathbf{k}_m}\rangle + b_i^{v\mathbf{k}_m} |\Phi^{val}_{b,i\mathbf{k}_m}\rangle$ and $|\psi_{c\mathbf{k}_m}\rangle = \sum_i a_i^{c\mathbf{k}_m} |\Phi^{cond}_{a,i\mathbf{k}_m}\rangle + b_i^{c\mathbf{k}_m} |\Phi^{cond}_{b,i\mathbf{k}_m}\rangle$, where $\Phi_a$ and $\Phi_b$ refer to the pristine superlattice wavefunctions of the bottom and top WS$_2$ layer, respectively. The pristine superlattice states are related to states in the rotated pristine BZs of the individual layers by band folding. Using the set of expansion coefficients, $a_i^{v\mathbf{k}_m}$, $b_i^{v\mathbf{k}_m}$, $a_i^{c\mathbf{k}_m}$ and $b_i^{c\mathbf{k}_m}$, we approximate each moiré electron-hole interaction kernel matrix element in the BSE as a coherent sum over many pristine unit-cell kernel matrix elements. This generalized PUMP method allows us to calculate moiré exciton energies and wavefunctions for the t-WS$_2$ moiré superlattice (see SI for more details). We study the $\mathbf{Q} = 0$ exciton in the moiré BZ, where $\mathbf{Q}$ is the exciton center-of-mass wave vector. While these states are formed by coherent superposition of direct transitions in the moiré BZ, the valence-band and conduction-band states involved originate from different valleys in the unfolded unit-cell BZ and are strongly modified by the moiré potential.

Fig. 3d shows the calculated electron density distribution for the lowest-energy exciton wavefunction $\chi_0(\mathbf{r}_e, \mathbf{r}_h)$ with a fixed hole position $\mathbf{r}_h$ (labeled with a red solid dot) at the AB site (left panel), the B$^{W/W}$ site (middle panel), and the B$^{S/S}$ site (right panel). The electron density is maximum at the B$^{W/W}$ site when the hole is at the AB site and is vanishingly small when the hole is at the B$^{W/W}$ and B$^{S/S}$ sites. The spin-aligned and spin-antialigned excitons have the same energy due to spatial separation of photoexcited electron and hole, which results in a small exchange energy. In the out-of-plane direction, the electron and hole charge densities are delocalized over both layers and show no interlayer charge transfer. This demonstrates that the lowest-energy exciton state in t-WS$_2$ is a layer-hybridized in-plane charge-transfer (ICT) exciton, in which the electron and hole prefer the B$^{W/W}$ site and AB site, respectively. The calculated binding energy for the ICT exciton, when the bilayer is suspended in vacuum, is ~150 meV. In presence of the hBN substrate, we expect the binding energy to be smaller due to additional dielectric screening.

ICT moiré excitons can be expected to yield a STM tip-position dependent photocurrent polarity due to their intrinsic lateral electron-hole separation. As illustrated in Fig. 3e, when the STM tip is parked above the photoexcited electron of an ICT exciton (left panel) the tunneling probability for the electron dominates compared to the spatially displaced photoexcited hole, thus resulting in negative tunneling current. Similarly, positive tunnel current is expected when the tip is parked above the photoexcited hole (right panel). Our photocurrent measurement thus directly reflects the charge density distribution of the lowest-energy ICT moiré exciton which can be expressed in unit of proton charge as $\rho(r) = \rho_h(r) - \rho_e(r)$, where $\rho_e(r) = \int |\chi_0(r_e, r_h)|^2 dr_h$ and $\rho_h(r) = \int |\chi_0(r_e, r_h)|^2 dr_e$ are the electron and hole densities, respectively. The calculated distribution map for $\rho(r)$ is shown in Fig. 3f. The laterally separated electrons and holes of the ICT excitons yield an alternating charge polarity that nicely matches our photocurrent map (Fig. 3c), providing quantitative spatial evidence that we are imaging ICT moiré excitons.

We are able to explore the interaction between the STM tip and ICT moiré excitons by investigating the $V_{bias}$ dependence of the photocurrent. Fig. 4a shows the photocurrent as a function of $V_{bias}$ measured at the AB (red, hole dominant) and $B^{W/W}$ (blue, electron dominant) sites. It shows a ~200mV tip bias range (marked by dashed lines) where positive photocurrent (at AB sites) and negative photocurrent (at $B^{W/W}$ sites) coexist. To observe the tip's effect on the photocurrent we measured photocurrent maps for -985mV ≤ $V_{bias}$ ≤ -602mV using the same tip (Fig.4b-4f). These maps also show a ~200mV range of $V_{bias}$ values where positive and negative photocurrent coexist, consistent with Fig. 4a (this $V_{bias}$ range varies depending on the tip sharpness but always lies between 100mV and 250mV (more $V_{bias}$ dependent photocurrent maps are shown in the SI)).

The $V_{bias}$ dependence of the photocurrent reflects a tip-induced exciton dissociation effect, as illustrated in Fig. 4g-4i for different $V_{bias}$ values. Because of capacitive coupling between the backgate graphite and the STM tip, a potential difference between them induces an electric field near the tip apex that perturbs the ICT moiré excitons (the t-WS$_2$ chemical potential lies within an energy gap at V$_{BG}$ = 0 and so does not screen the electric field). The work function difference between the tip (made of Pt/Ir) and the graphite back gate causes that this field to be present even at V$_{bias}$ = 0, but it can be cancelled by setting V$_{bias}$ = V$_0$ where V$_0$ is the work function difference. At $V_{bias} - V_0 = 0$ (Fig. 4g) the tip exerts only a weak pertubation on the ICT excitons and so a spatially alternating photocurrent polarity can be observed (Fig. 3c and Fig. 4c-4e). However, at $V_{bias} - V_0 > 0$ the tip apex accumulates negative charge and attracts holes while repelling electrons (Fig. 4h). Above a certain threshold this effect dissociates the ICT excitons and only a positive current can be observed as in Fig. 4f. Similarly, only a negative current is seen when $V_{bias} - V_0$ is much smaller than zero (Fig. 4b) due to opposite dissociation of the ICT excitons, as illustrated in Fig. 4i. The experimentally observed V$_0$ ranges from -800mV to -300mV depending on the tip structure (V$_0$ = -790mV for the measurements shown in Fig. 4) (further details of the tip-exciton interaction are discussed in the SI).

In conclusion, our photocurrent tunneling microscopy enables real-space imaging of in-plane charge-transfer moiré excitons with sub-nanometer spatial resolution. The observed electron and hole distributions agree well with *ab initio GW*-BSE calculation results. This work establishes a new approach for probing the microscopic behavior of photoexcited states in 2D van der Waals heterostructures.

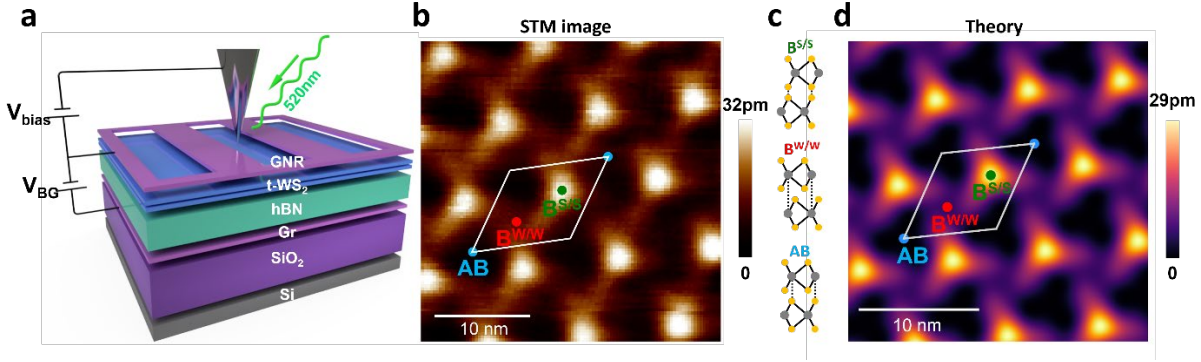

**Figure 1. Laser-STM measurement of a twisted bilayer WS$_2$ moiré superlattice. a.** Sketch of experimental setup for laser-STM measurement of a near-58-degree twisted bilayer WS$_2$ (t-WS$_2$) device. The t-WS$_2$ is placed on top of 49nm thick hBN and a graphite substrate (these serve as the gate dielectric and back gate). A back gate voltage $V_{BG}$ is applied between the t-WS$_2$ and the graphite back gate. A graphene nanoribbon (GNR) array is placed on top of the t-WS$_2$ to serve as the contact electrode. A sample-tip bias $V_{bias}$ is applied between the t-WS$_2$ and the STM tip to induce a tunnel current. A 520nm wavelength continuous-wave laser is focused on the tip tunnel junction. **b.** A typical STM topography image of the t-WS$_2$ surface exhibits the triangular moiré lattice. $V_{bias}$ = -4V, I = 100pA. Three high-symmetry stacking regions are labeled with solid dots: $B^{S/S}$, $B^{W/W}$, and AB. **c.** Illustration of the $B^{S/S}$, $B^{W/W}$, and AB stacking structures (yellow and gray dots represent S and W atoms respectively). **d.** Theoretically calculated surface height variation for t-WS$_2$.

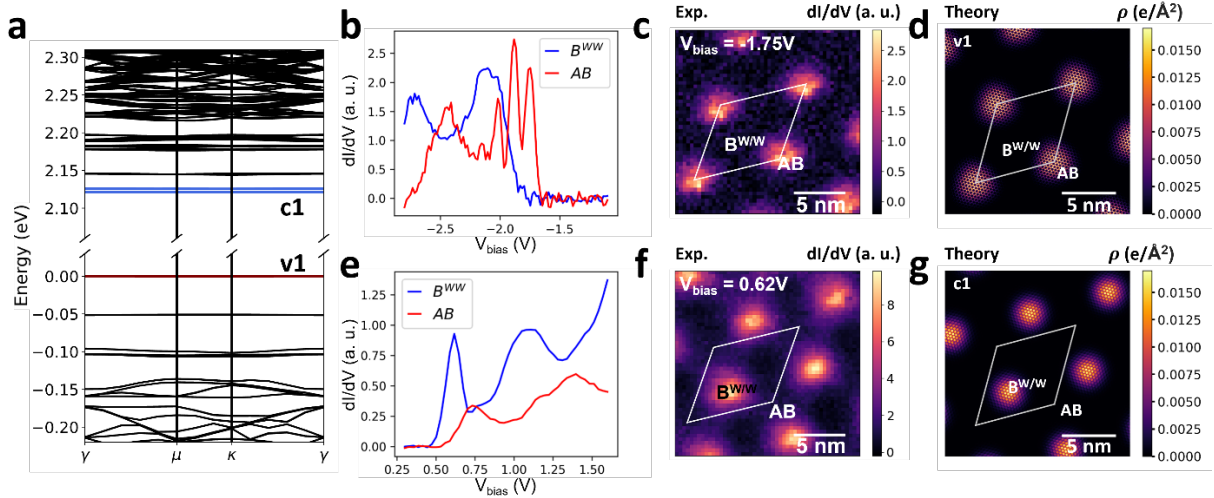

**Figure 2. Electronic structure of twisted bilayer WS$_2$. a**. Calculated electronic band structure for the t-WS$_2$ moiré superlattice. The moiré flat bands at the valence band top (VBT) and conduction band bottom (CBB) are labeled v1 and c1. **b.** dI/dV spectra measured at the $B^{W/W}$ (blue) and AB (red) stacking sites for the valence band. The AB site shows a first sharp peak near $V_{bias}$ = -1.75V. (**c**) t-WS$_2$ dI/dV map measured at $V_{bias}$ = -1.75 V and (**d**) calculated charge density ($\rho$) distribution for v1 both show strong hole localization at the AB site. $V_{BG}$ = -2V for (**b,c**). **e**. dI/dV spectra measured at the $B^{W/W}$ (blue) and AB (red) stacking sites for the conduction band. The $B^{W/W}$ site shows a first sharp peak near $V_{bias}$ = 0.60V. (**f**) t-WS$_2$ dI/dV map

measured at $V_{bias} = 0.62V$ and (**g**) calculated charge density ($\rho$) distribution for c1 both show strong electron localization at the $B^{W/W}$ site. $V_{BG} = 1.5V$ for (**e,f**).

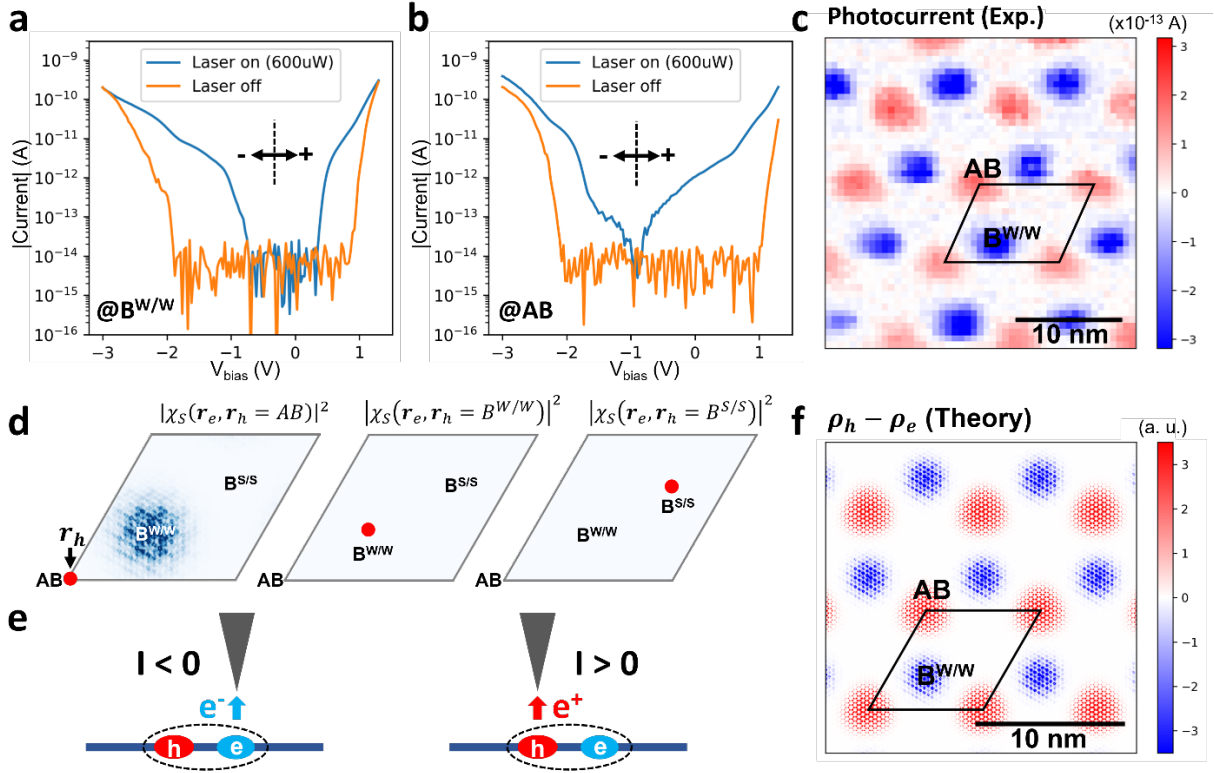

**Figure 3. Photocurrent mapping of in-plane charge transfer excitons. a, b**. STM tunnel current spectra measured at the (**a**) $B^{W/W}$ and (**b**) AB stacking sites with the laser turned off (orange) and on (blue). $V_{BG} = 0$. The absolute value of the current is plotted on a logarithmic scale (the left and right branch of the spectra correspond to negative and positive current respectively). For the laser-off case, the current at both the $B^{W/W}$ and the AB sites reflect an energy gap for $-2V < V_{bias} < 1V$. For the laser-on case (P = 600uW) photocurrent emerges in the energy gap region and the $B^{W/W}$ and AB sites show different photocurrent spectral shapes. **C**. A photocurrent map of t-WS$_2$ measured with the laser on (P = 600uW) for $V_{bias} = -0.60V$ and $V_{BG} = 0$ shows positive (negative) photocurrent at the AB ($B^{W/W}$) sites. **d**. Calculated electron density for the lowest-energy exciton, $|\chi_0(r_e, r_h)|^2$, with fixed hole position $r_h$ (labeled with red solid dot) at the AB site (left panel), the $B^{W/W}$ site (middle panel), and the $B^{S/S}$ site (right panel). The maps show appreciable electron density at the $B^{W/W}$ site only when the hole position is fixed at the AB site (left panel). **e**. Schematic for tip-position dependent tunnel current from an ICT exciton. When the STM tip sits above the electron (left panel) the larger tunnel probability for the electron yields a negative current. A positive current is detected when the tip sits above the hole (right panel). **f**. Calculated charge distribution map $\rho(r) = \rho_h(r) - \rho_e(r)$ of the ICT exciton, where $\rho_h(r) = \int |\chi_0(r_e, r_h)|^2 dr_e$ and $\rho_e(r) = \int |\chi_0(r_e, r_h)|^2 dr_h$.

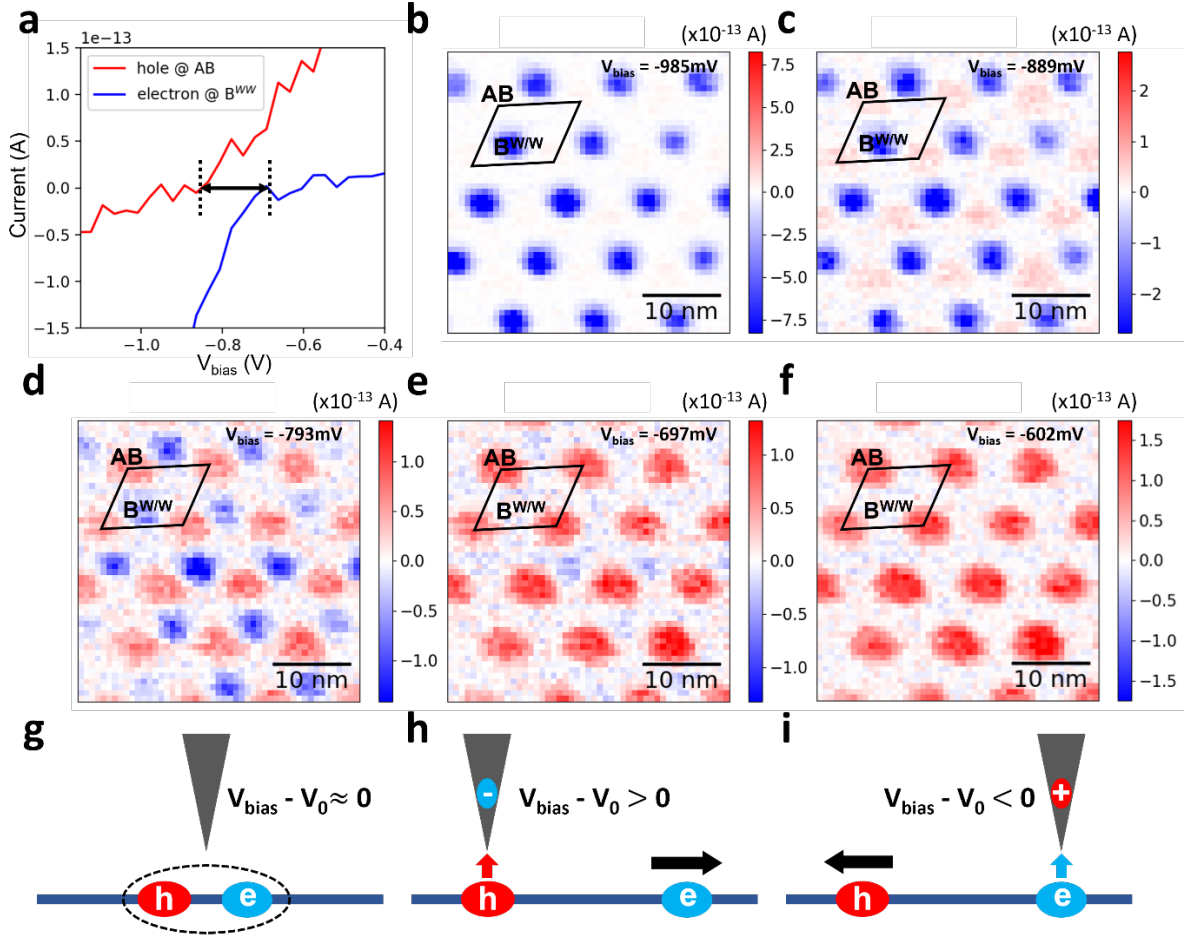

**Figure 4. Interaction between STM tip and ICT excitons. a.** Zoom-in photocurrent spectra at the AB (hole) site (red) and the $B^{W/W}$ (electron) site (blue). Spatially alternating current polarity occurs for a $V_{bias}$ range of ~200mV (dashed line). **b-f**. Evolution of t-WS$_2$ photocurrent maps for increasing $V_{bias}$: **(b)** $V_{bias}$ = -985mV, **(c)** $V_{bias}$ = -889mV, **(d)** $V_{bias}$ = -793mV, **(e)** $V_{bias}$ = -697mV, and **(f)** $V_{bias}$ = -602mV. Spatially alternating current polarity exists only in **c-e**, while negative (positive) current dominates in **b** (**f**). **g-i**. Diagram of tip-induced ICT exciton dissociation effect. $V_0$ is the bias voltage offset that compensates the work function difference between the tip and the back gate graphite. **g**. For $V_{bias} - V_0 \approx 0$ the tip does not significantly perturb the ICT exciton and so both photocurrent polarities are seen. **h.** For $V_{bias} - V_0 > 0$ negative charge accumulates at the tip apex which attracts holes and repels electrons, thereby dissociating ICT excitons. **i**. For $V_{bias} - V_0 < 0$ the tip attracts electrons and repels holes, thereby dissociating ICT excitons.

**Corresponding Author**

\* Email: sglouie@berkeley.edu (S.L.), crommie@physics.berkeley.edu (M.C.) and fengwang76@berkeley.edu (F.W.).13

**Author Contributions**

S.G.L., M.C., and F.W. conceived the project. H.L. and Z.X. performed the STM/STS and PTM measurement, M.N. and F.H.J. formulated the generalized PUMP method, M.N., W.K. and Z.L. performed the *ab initio GW*-BSE calculations. H.L. and Z.X. fabricated the heterostructure device. R.S., R.B. and S.T. grew the $WS_2$ crystals. K.W. and T.T. grew the hBN single crystal. All authors discussed the results and wrote the manuscript.

**Notes**

The authors declare no financial competing interests.

**ACKNOWLEDGMENT**

This work was primarily funded by the U.S. Department of Energy, Office of Science, Office of Basic Energy Sciences, Materials Sciences and Engineering Division under Contract No. DE-AC02-05-CH11231 (van der Waals heterostructure program KCFW16) (device fabrication, STM spectroscopy, and force-field calculations for structural reconstructions). The Center for Computational Study of Excited-State Phenomena in Energy Materials (C2SEPEM) at Lawrence Berkeley National Laboratory, supported by the US Department of Energy, Office of Science, Basic Energy Sciences, Materials Sciences and Engineering Division under contract no. DE-AC02-05CH11231, as part of the Computational Materials Sciences Program provided advanced codes and experimental support of optical measurements. The Theory of Materials Program (KC2301) funded by the DOE Office of Science, Basic Energy Sciences, Materials Sciences and Engineering Division under contract DE-AC02-05CH11231, provided resources to develop the PUMP approach and analysis of the moiré excitons. Computational resources were provided by National Energy Research Scientific Computing Center (NERSC), which is supported by the DOE Office of Science under contract DE-AC02-05CH11231, and Frontera at TACC, which is supported by the National Science Foundation under grant OAC-1818253. Support was also provided by the National Science Foundation Award DMR-2221750 (surface preparation). S.T acknowledges support from DOE-SC0020653, NSF DMR 2111812, NSF DMR 1552220, NSF 2052527, DMR 1904716, and NSF CMMI 1933214 for $WS_2$ bulk crystal growth and analysis. K.W. and T.T. acknowledge support from JSPS KAKENHI (Grant Numbers 19H05790, 20H00354 and 21H05233). We thank Y. W. Choi, S. Kundu, and J. Ruan for discussions.

**Methods**

**Sample fabrication:** The twisted $WS_2$ device was fabricated using a micromechanical stacking technique[46]. A poly(propylene) carbonate (PPC) film stamp was used to pick up all exfoliated 2D material flakes. The 2D material layers in the main heterostructure region were picked up in the following order: substrate hBN, graphite, bottom hBN, monolayer $WS_2$, twisted monolayer $WS_2$, graphene nanoribbon array. The graphene nanoribbon array serves as a contact electrode for the twisted $WS_2$. The PPC film, together with the stacked sample, was peeled, flipped over, and transferred onto a $Si/SiO_2$ substrate ($SiO_2$ thickness 285nm). The PPC layer was subsequently removed using ultrahigh vacuum annealing at 330 °C, resulting in an atomically-clean heterostructure suitable for STM measurements. A 50nm Au and 5nm Cr metal layer was evaporated through a shadow mask to form electrical contacts to graphene layers.

**dI/dV spectroscopy measurement**: A bias modulation having 25mV amplitude and 500~900 Hz frequency was used to obtain the dI/dV signal. All the dI/dV spectra and mapping in Fig.2 were measured at T=5.4K. dI/dV mapping was performed under open-loop conditions with the tip height set by the following procedure: (a) set tip height at the condition $V_{bias}$ = 3.8V and I = 10 pA with closed feedback, (b) turn off the feedback and lower the tip height by a distance of $h_{tip}$ = 50pm.

**Auto laser alignment**: In order to overcome laser misalignment induced by STM thermal drift during the PTM measurement (usually lasting ~10 hours), a home-built auto laser alignment system with feedback control was developed. The laser spot position is detected with camera and corrected via a piezo-driven mirror every 20s.

**PTM measurement**: All the photocurrent tunneling spectra and mapping shown in this work were measured at T=7.6K~7.7K with the laser power ranging from 600uW to 900uW. The photocurrent mapping was performed under open-loop conditions with the tip height set by the following procedures: (a) set tip height at tip bias $V_{bias}$ and current setpoint I, (b) turn off feedback control and lower tip height by distance $h_{tip}$. Parameters for Fig.3c: $V_{bias}$ = -3V, I = 400pA, and $h_{tip}$ = 0. Parameters for Fig. 4b-4f: $V_{bias}$ = -3V, I = 200pA, and $h_{tip}$ = 0. The parameters used for extended photocurrent map data are specified in the SI.

**Theoretical calculations**: We studied a 57.72° twisted bilayer $WS_2$ moiré superlattice which contained 3786 atoms in the unit-cell. The structure is relaxed using classical forcefield calculations. The relaxed configuration of atoms is used to study the electronic structure of the moiré superlattice. We constructed the BSE effective Hamiltonian using 12 valence and 50 conduction bands with a $3 \times 3 \times 1$ k-point sampling of the moiré BZ. We used a greater number of conduction band states than valence band states due to the larger density of states at the conduction band edge (Fig. 2a). The moiré kernel matrix elements of the BSE were computed using the generalized PUMP method as a coherent linear combination of pristine unit-cell matrix elements (see SI for details). We used a basis of 200 valence and 200 conductions states of each pristine layer to expand the moiré electronic wavefunctions. The static-dielectric matrix used to compute the pristine unit-cell kernel matrix elements is approximated to be identical to that of pristine 2H or AB stacking.

**Supplementary Materials**

**Data availability**

The data supporting the findings of this study are included in the main text and in the Supplementary Information files, and are also available from the corresponding authors upon request.

# Supplementary Information for

# Imaging Moiré Excited States with Photocurrent Tunneling Microscopy

*Hongyuan Li*[1, 2, 3, 10], *Ziyu Xiang*[1, 2, 3, 10], *Mit H. Naik*[1, 3, 10], *Woochang Kim*[1, 3], *Zhenglu Li*[1, 3], *Renee Sailus*[4], *Rounak Banerjee*[4], *Takashi Taniguchi*[5], *Kenji Watanabe*[6], *Sefaattin Tongay*[4], *Alex Zettl*[1, 3, 7], *Felipe H. da Jornada*[8, 9], *Steven G. Louie*[1, 3]\*, *Michael F. Crommie*[1, 3, 7]\* and *Feng Wang*[1, 3, 7]\*

[1]Department of Physics, University of California at Berkeley, Berkeley, CA, USA.

[2]Graduate Group in Applied Science and Technology, University of California at Berkeley, Berkeley, CA, USA.

[3]Materials Sciences Division, Lawrence Berkeley National Laboratory, Berkeley, CA, USA.

[4]School for Engineering of Matter, Transport and Energy, Arizona State University, Tempe, AZ, USA.

[5]International Center for Materials Nanoarchitectonics, National Institute for Materials Science, Tsukuba, Japan

[6]Research Center for Functional Materials, National Institute for Materials Science, Tsukuba, Japan

[7]Kavli Energy Nano Sciences Institute at the University of California Berkeley and the Lawrence Berkeley National Laboratory, Berkeley, CA, USA.

[8]Department of Materials Science and Engineering, Stanford University, Palo Alto, CA, USA.

[9]Stanford Institute for Materials and Energy Sciences, SLAC National Accelerator Laboratory, Menlo Park, CA, USA.

[10]These authors contributed equally: Hongyuan Li, Ziyu Xiang, and Mit H. Naik

## S1. Images of the laser-STM setup

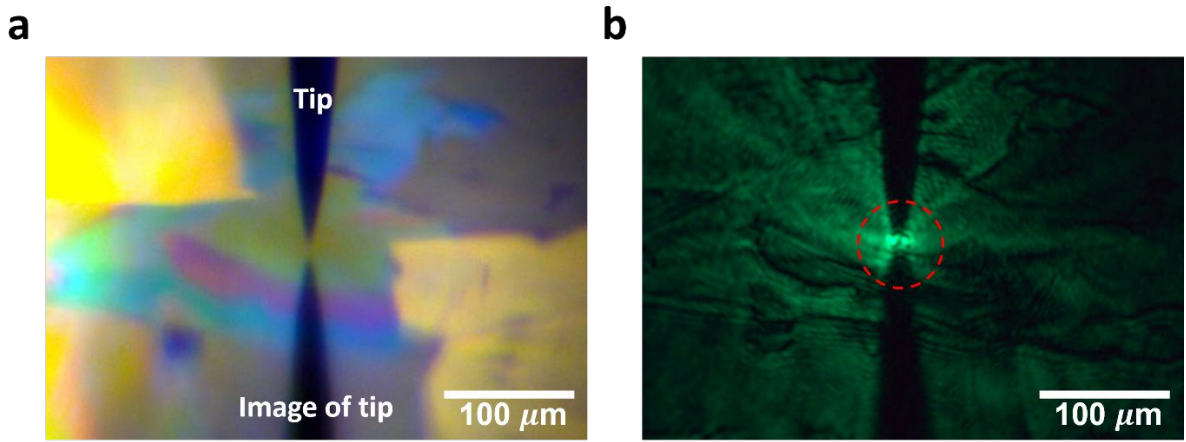

**Figure S1. Optical images of the tip-sample tunneling junction. a.** Optical image for gated twisted $WS_2$ device and laser-STM setup using white light illumination. The positions of the tip and its reflected image by the substrate are labeled. **b**. Optical image of the device and setup with a 520nm continuous-wave laser focused onto the tip-sample tunneling junction. The laser spot (diameter ~10$\mu$m) is labeled with a red dashed-line circle.

## S2. Extended data on the V$_{bias}$ dependence of the ICT exciton photocurrent maps

Fig. S2~S4 show extended data on the V$_{bias}$ dependent photocurrent maps of the ICT moire excitons in t-WS$_2$ for different tip status and scanning conditions.

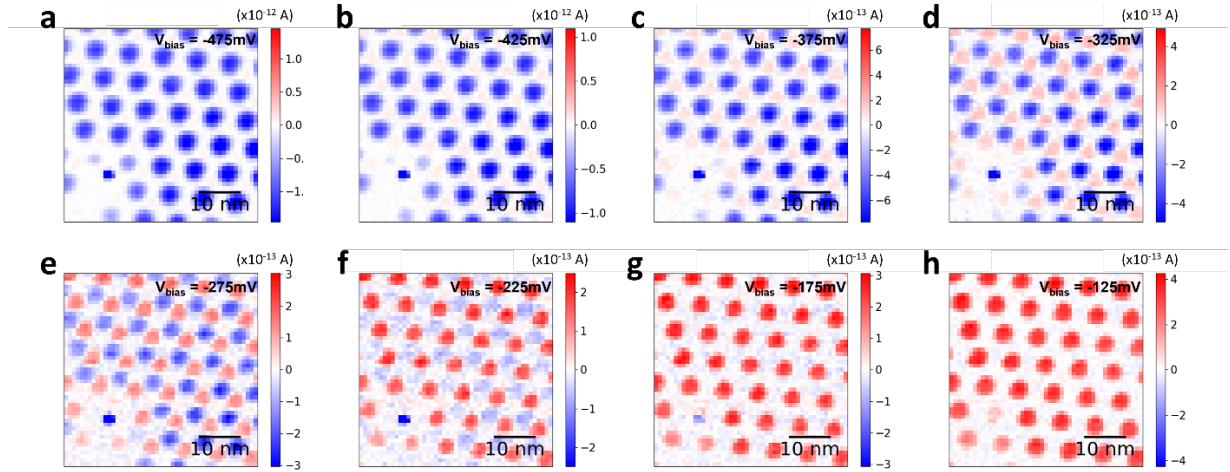

**Figure S2 Extended data on the V$_{bias}$-dependence of the ICT exciton photocurrent maps.** V$_{BG}$ = -0.3V, V$_{bias}$ = -3.4V, I = 11pA, h$_{tip}$ = 200pm, and P = 900uW. STM bias offset V$_0$ = -275mV. V$_{bias}$ range for coexisting positive and negative photocurrent: -375mV ≤ V$_{bias}$ ≤ -225mV.

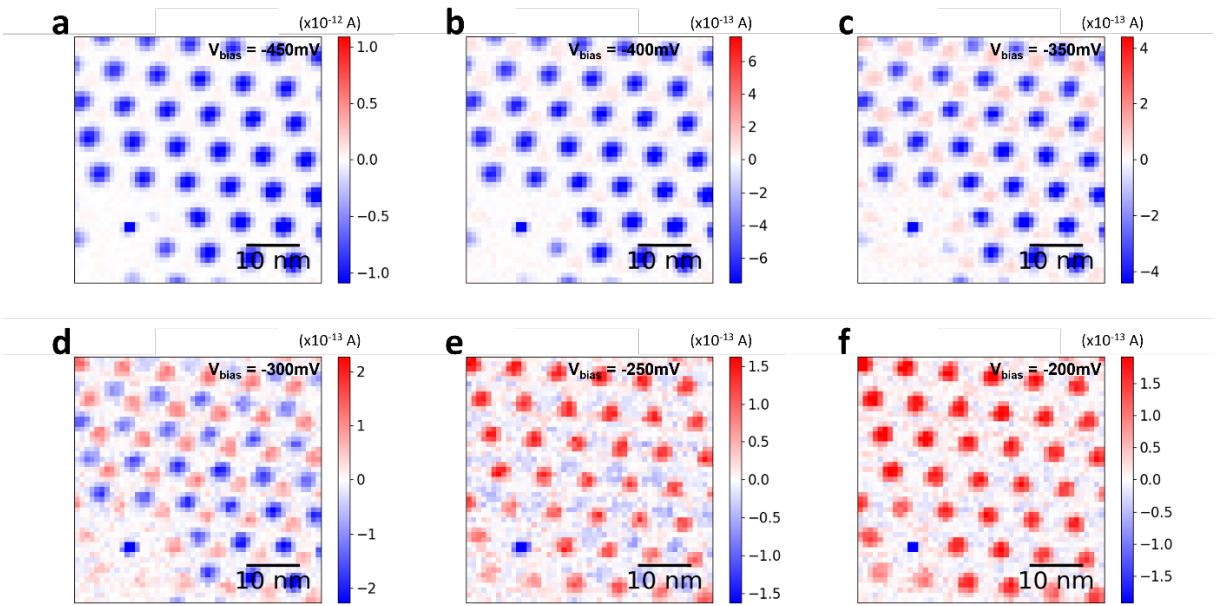

**Figure S3 Extended data on the V$_{bias}$-dependence of the ICT exciton photocurrent maps.**

V$_{BG}$ = -0.1V, V$_{bias}$ = -3.4V, I = 11pA, h$_{tip}$ = 160pm, and P= 900uW. STM bias offset V$_0$ = -300mV. V$_{bias}$ range for coexisting positive and negative photocurrent: -350mV ≤ V$_{bias}$ ≤ -250mV.

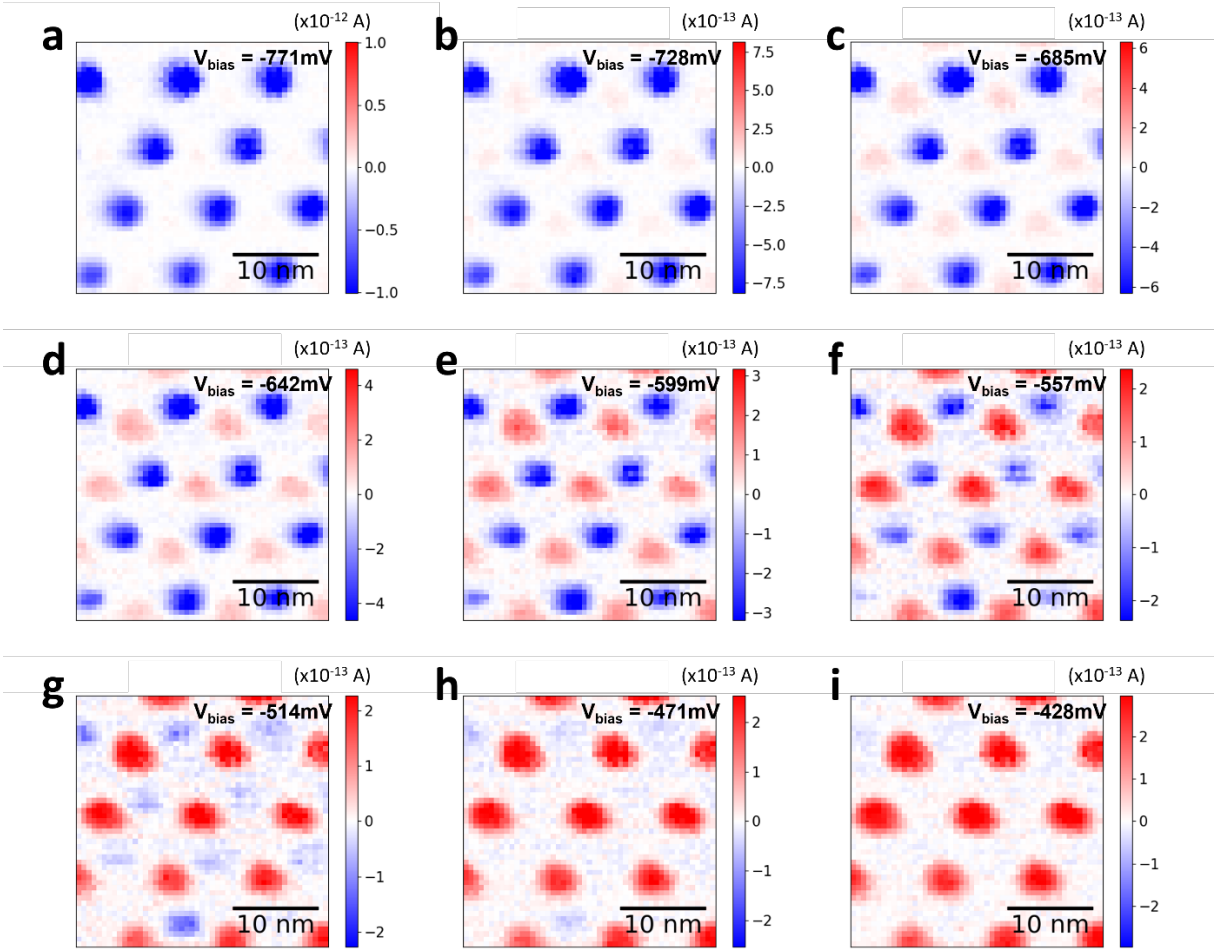

**Figure S4 Extended data on the V$_{bias}$-dependence of the ICT exciton photocurrent maps.**

V$_{BG}$ = 0, V$_{bias}$ = -3 V, I = 400pA, h$_{tip}$ = 0, and P= 600uW. STM bias offset V$_0$ = -599mV. V$_{bias}$ range for coexisting positive and negative photocurrent: -685mV ≤ V$_{bias}$ ≤ -471mV.

## S3. Ideal tunneling of ICT exciton without STM tip perturbation

Fig. S5a (I-III) illustrates the band diagram for the ideal tunneling of ICT moire excitons for different $V_{bias}$ conditions controlling the difference between the chemical potentials of the t-WS$_2$ ($\mu_{tWS_2}$) and the tip ($\mu_{tip}$). When $\mu_{tip}$ is within the bandgap of the t-WS$_2$ (regime I), both the electron and hole of the ICT exciton can tunnel into the STM tip (with probabilities depending on the tip position). When $V_{bias}$ is more negative such that $\mu_{tip}$ is below the energy of the exciton's hole (labeled with a red peak) (regime II), the tunneling of the hole is forbidden due to the Pauli exclusion principle. Similarly, the tunneling of the electron into the tip is forbidden when $\mu_{tip}$ is above the energy of the exciton's electron (regime III). Fig. S5b illustrates the ideal tunnel photocurrent as a function of $V_{bias}$ at B$^{W/W}$ sites and AB sites where electrons and holes localize, respectively. Negative tunnel current due to electron tunneling exists when $eV_{bias} > -(E_g - E_b)$ while positive tunnel current due to hole tunneling exists when $eV_{bias} < (E_g - E_b)$, where $E_g$ (~1.3eV) and $E_b$ (~20meV) are the band gap and ICT exciton binding energy, respectively.

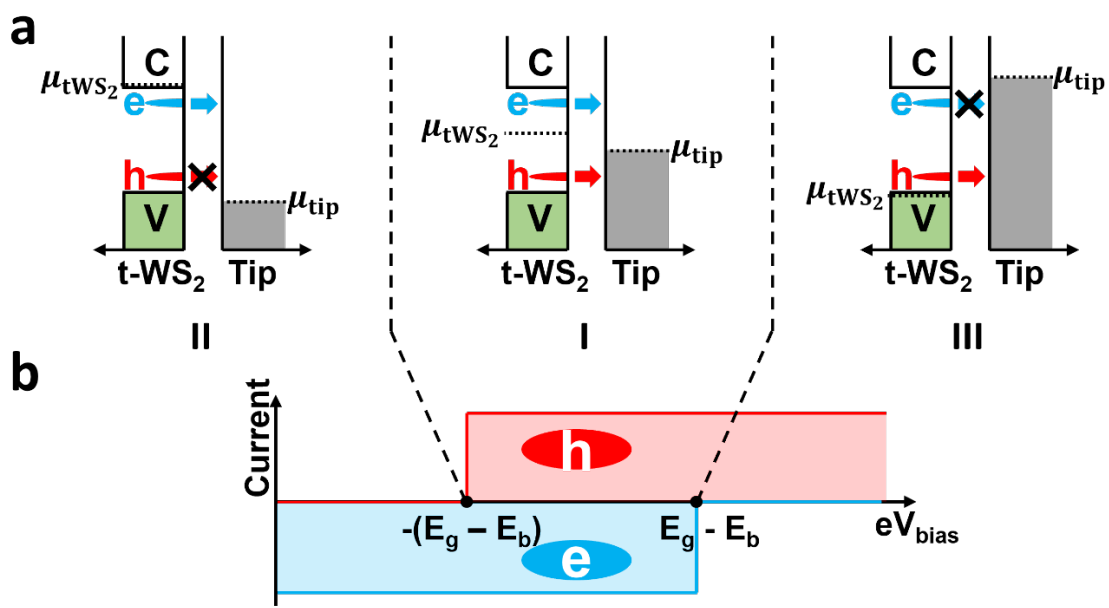

**Figure S5. Ideal tunneling of ICT exciton without STM tip perturbation for different $V_{bias}$ conditions.** See full discussion in S3.

## S4. Structural reconstruction of the moiré superlattice (Theory)

We modeled a moiré superlattice using two monolayers of $WS_2$ by applying a twist-angle of 57.72°. The monolayer unit-cell lattice constant was chosen to be 3.15 Å. The structure was constructed and relaxed using the TWISTER[1] code. The moiré unit-cell contained 3786 atoms. The structural relaxation of the moiré superlattice was performed using classical force-field calculations as implemented in the LAMMPS[2] package. The intralayer forces in the $WS_2$ layers were modeled using the Stillinger-Weber[3] force-field, and the interlayer interactions are included using a registry-dependent Kolmogorov-Crespi[4,5] force-field. The lowest-energy stacking configuration, AB or 2H, had the largest area in the reconstructed moiré superlattice[5–7]. The interlayer spacing between the layers was also modulated in the superlattice, with the $B^{S/S}$ stacking having the largest spacing due to steric hindrance between the S atoms of the two layers. The variation of interlayer spacing between the layers led to an inhomogeneous interlayer hybridization[6,8] in the moiré superlattice and influences the electronic structure. The reconstruction also led to a strain redistribution in the individual layers. Fig. S6 shows the interlayer spacing distribution and strain reconstruction in the 57.72° twisted bilayer $WS_2$. The strain was localized at the domain boundary between the AB and $B^{W/W}$ stacking[7].

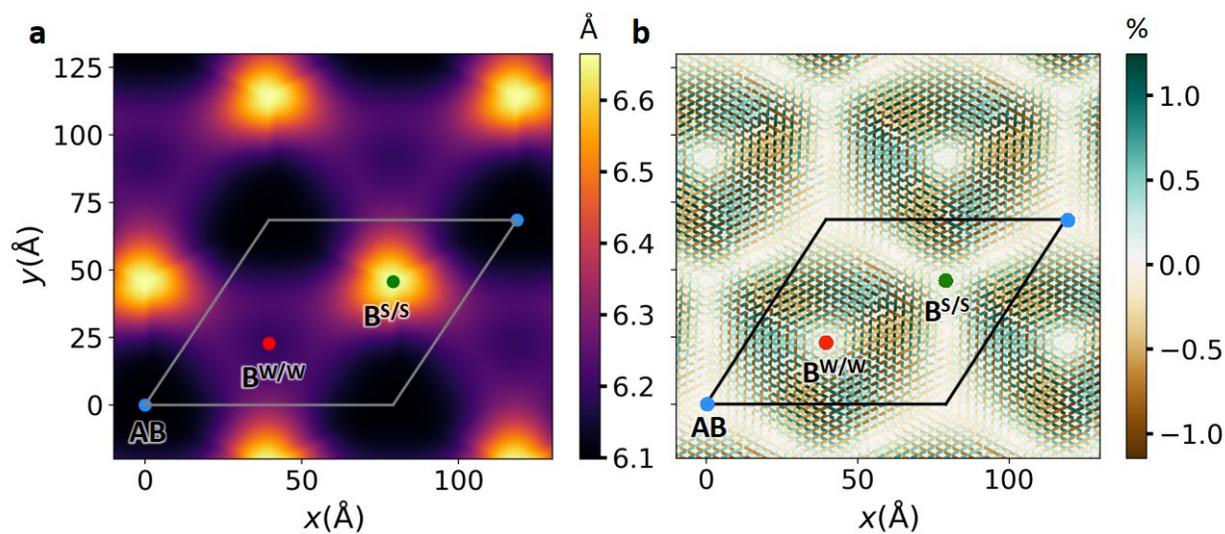

**Figure S6. Theoretical structural reconstruction of the moiré superlattice. a.** Distribution of the interlayer spacing (calculated as the local distance between the W atoms) in the 57.72° twisted bilayer $WS_2$ moiré superlattice. **b.** Strain redistribution in the bottom layer of the moiré superlattice. The strain is plotted as the percentage change in the local W-W distance from the pristine monolayer lattice constant.

## S5. Electronic structure and optical calculations

The density functional theory (DFT) electronic structure calculations were carried out starting with the relaxed structure of the superlattice obtained from the force-field calculations described above. The calculations were performed using the SIESTA[9,10] code which uses localized atomic orbitals as the basis set, making it is more efficient for studying moiré unit-cells containing thousands of atoms compared to a DFT implementation using plane-wave basis sets. Relativistic optimized norm-conserving Vanderbilt (ONCV) pseudopotentials[11] used in the calculation were obtained from the PseudoDojo repository[12,13], and the exchange-correlation functional was approximated using the generalized gradient approximation[14]. Spin-orbit interaction was included in all electronic structure calculations. To capture hybridization between layers to the same level of accuracy as plane-wave basis sets, we expanded[15] the atomic basis set of the S atom to also include the 4s and 4p orbitals which have a larger spatial extent. We only sampled the zone-center $\gamma$ point to obtain the self-consistent charge density. All the pristine unit-cell calculations were performed using the Quantum Espresso[16] code, which uses plane-wave basis sets. The electronic states were expanded in plane waves up to an energy cutoff of 40 Ry. We ensured that the same pseudopotential and exchange-correlation functional were used in the SIESTA and Quantum Espresso calculations. We found that the electronic states calculated using SIESTA have an overlap of over 98% with the corresponding states calculated using Quantum Espresso. We tested this for valence and conduction band edge states of pristine unit-cell AB stacked bilayer $WS_2$ for all the k-points in a $12 \times 12 \times 1$ sampling of the Brillouin zone (BZ).

The electronic band structure and the charge density of states close to the valence and conduction band edge are plotted in Fig. S7 for the 57.72° twisted bilayer $WS_2$. The band gap at the DFT level is severely underestimated, while the dispersions of the bands is relatively well captured. We computed the quasiparticle band gap for a 2H stacked bilayer by computing the electron self-energy at the $GW$ level[17]. We used the $GW$ correction to the $\Gamma - Q$ band gap to correct the DFT band structure of the moiré superlattice. The $GW$ calculations were performed using the BerkeleyGW[18] package. The random phase approximation was used to compute the static dielectric matrix and screened Coulomb interaction. The plane-wave cutoff for the dielectric matrix was 30 Ry and ~6000 unoccupied states were employed. The Hybertsen-Louie generalized plasmon pole model[17] was used to extend the static dielectric to finite frequencies. The convergence of the band gap with q-point sampling was greatly improved[19] by using the recently developed nonuniform neck subsampling[20] method. The Coulomb interaction was also truncated in the out-of-plane direction[21].

Inhomogeneous interlayer hybridization and strain reconstruction leads to a deep moiré potential[7] that confines the valence band edge states to the AB site and conduction band edge states to the $B^{W/W}$ site in 57.72° twisted bilayer $WS_2$ (Fig. S7). The spatial distribution of the states at the valence band edge closely resembles low-energy states of a particle in an ideal triangular quantum well potential[7]. These findings are in good agreement with previous electronic structure calculations[7] on twisted bilayer $MoS_2$.

The electron-hole interaction kernel matrix elements of the Bethe-Salpeter equation (BSE) were computed using the BerkeleyGW package[18,22]. The pristine unit-cell matrix projection (PUMP) method (described in detail below) was used to compute the moiré superlattice kernel matrix elements as a linear combination of many pristine unit-cell kernel matrix elements. For the 57.72° twisted bilayer WS$_2$ we constructed the BSE effective Hamiltonian using 12 valence and 50 conduction bands with moiré BZ k-point sampling of $3 \times 3 \times 1$. We used a greater number of conduction band states than valence band states due to the larger density of states at the conduction band edge (Fig. S7). The valence and conduction bands included in the BSE span an energy range of ~200 meV, which is larger than the binding energy of the in-plane charge transfer exciton. A static-dielectric matrix, calculated in the pristine unit-cell BZ for the AB stacking, was used to compute the pristine unit-cell kernel matrix elements. The dielectric matrix was calculated using the random phase approximation with a planewave energy cut-off of 3.5 Ry and ~400 empty states.

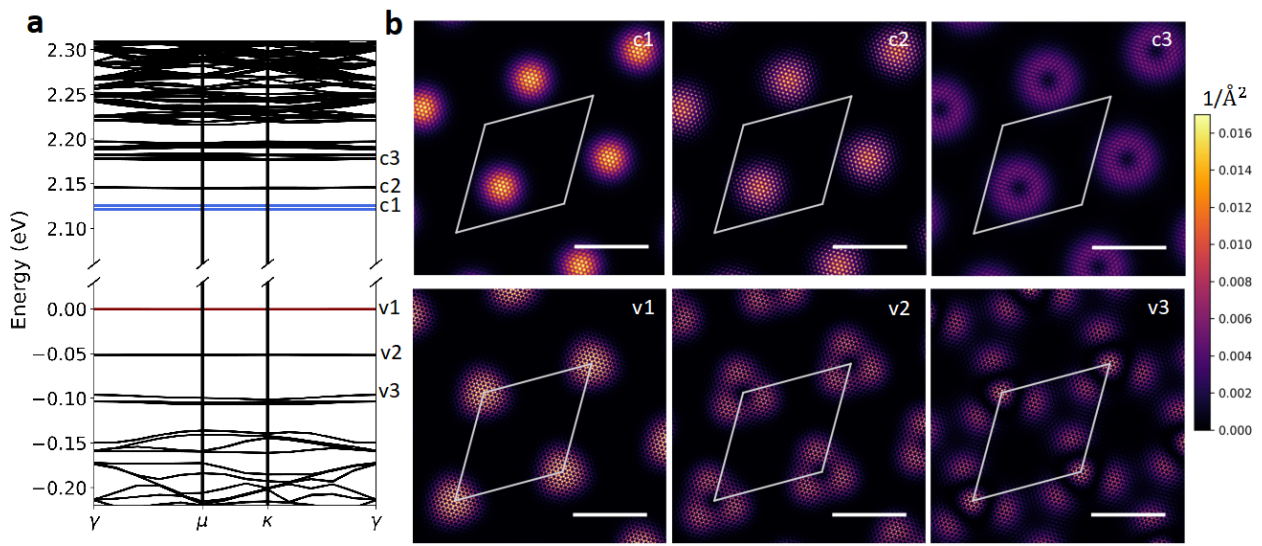

**Figure S7. Electronic band structure and wavefunction modulation in 57.72° twisted bilayer WS$_2$. a.** Electronic band structure of the 57.72° twisted bilayer WS$_2$ moiré superlattice including self-energy corrections at the *GW* level. **b.** Charge density distribution of the states at the $\gamma$ point in the moiré BZ as marked in **a.** The charge density is integrated along the out-of-plane direction.

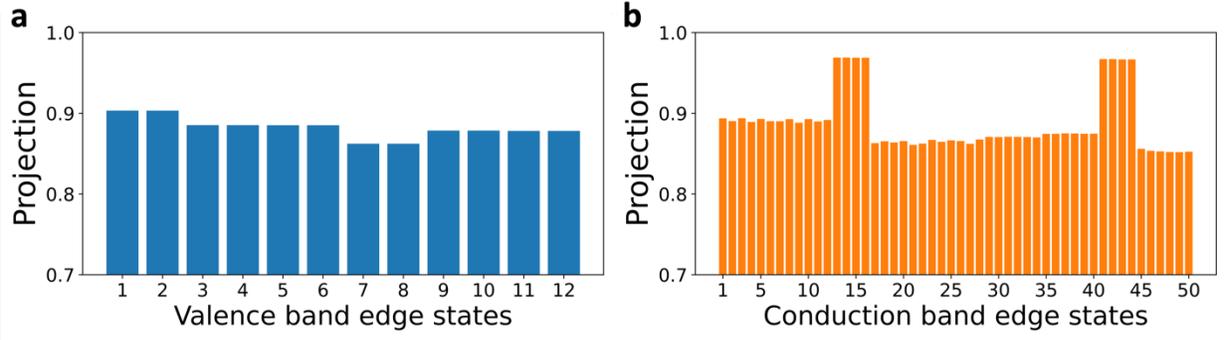

**Figure S8. Projection of the constructed wavefunctions on the original wavefunctions. a.** and **b.** Projection of the constructed 57.72° twisted bilayer WS$_2$ moiré superlattice valence ($\langle\psi^{cons}_{vk_m}|\psi_{vk_m}\rangle$) and conduction ($\langle\psi^{cons}_{ck_m}|\psi_{ck_m}\rangle$) electronic states on the original electronic states at the $\gamma$ point in the moiré BZ, respectively. The moiré electronic states are constructed using a basis of 200 valence and 200 conductions states of each pristine layer ($n = 200$ in Eq. S1). The average projection is 89%. The conduction states derived from the $K$ point in the pristine unit-cell BZ, which have less interlayer hybridization, have a higher projection compared to the states derived from the $Q$ point.

## S6. Generalized pristine unit-cell matrix projection method for GW-BSE calculations

In the generalized pristine unit-cell matrix projection (PUMP) approach, we first express the twisted bilayer WS$_2$ moiré superlattice electronic wavefunctions as a linear combination of pristine wavefunctions of the individual layers. We denote the bottom pristine layer as 'a' and top layer as 'b'. The moiré valence and conduction states are constructed from the pristine wavefunctions,

$$|\psi_{v\mathbf{k}_m}^{\text{cons}}\rangle = \sum_{i=1}^{n}(a_i^{v\mathbf{k}_m}|\Phi_{a,i\mathbf{k}_m}^{\text{val}}\rangle + b_i^{v\mathbf{k}_m}|\Phi_{b,i\mathbf{k}_m}^{\text{val}}\rangle), \quad (S1)$$

$$|\psi_{c\mathbf{k}_m}^{\text{cons}}\rangle = \sum_{i=1}^{n}(a_i^{c\mathbf{k}_m}|\Phi_{a,i\mathbf{k}_m}^{\text{cond}}\rangle + b_i^{c\mathbf{k}_m}|\Phi_{b,i\mathbf{k}_m}^{\text{cond}}\rangle). \quad (S2)$$

Here, $|\Phi_{a,i\mathbf{k}_m}^{\text{val}}\rangle$ and $|\Phi_{b,i\mathbf{k}_m}^{\text{val}}\rangle$ refers to pristine superlattice valence states of layer 'a' and layer 'b', respectively, and $|\Phi_{a,i\mathbf{k}_m}^{\text{cond}}\rangle$ and $|\Phi_{b,i\mathbf{k}_m}^{\text{cond}}\rangle$ to pristine superlattice conduction states of layer 'a' and layer 'b', respectively. We have verified that the new basis sufficiently describes the original moiré electronic wavefunctions ($|\psi_{v\mathbf{k}_m}\rangle$ and $|\psi_{c\mathbf{k}_m}\rangle$) by computing the overlaps $\langle\psi_{v\mathbf{k}_m}^{\text{cons}}|\psi_{v\mathbf{k}_m}\rangle$ and $\langle\psi_{c\mathbf{k}_m}^{\text{cons}}|\psi_{c\mathbf{k}_m}\rangle$ (see Fig. S8). Using the expansion coefficients, the BSE electron-hole interaction kernel matrix elements of the moiré superlattice can be approximated as a linear combination of pristine unit-cell kernel matrix elements of layer 'a' and layer 'b',

$$\langle\psi_{v\mathbf{k}_m}\psi_{c\mathbf{k}_m}|K|\psi_{v\mathbf{k}_m'}\psi_{c\mathbf{k}_m'}\rangle$$

$$\approx \sum_{\alpha\beta\gamma\lambda=a,b}\sum_{ijpq}\alpha_i^{v\mathbf{k}_m*}\beta_j^{c\mathbf{k}_m*}\gamma_p^{v\mathbf{k}_m}\lambda_q^{c\mathbf{k}_m}\langle\Phi_{\alpha,i\mathbf{k}_m}^{\text{val}}\Phi_{\beta,j\mathbf{k}_m}^{\text{cond}}|K|\Phi_{\gamma,p\mathbf{k}_m'}^{\text{val}}\Phi_{\lambda,q\mathbf{k}_m'}^{\text{cond}}\rangle \quad (S3)$$

$$= \sum_{\alpha\beta\gamma\lambda=a,b} \sum_{ijpq} \alpha_i^{v\mathbf{k}_m *} \beta_j^{c\mathbf{k}_m *} \gamma_p^{v\mathbf{k}_m} \lambda_q^{c\mathbf{k}_m} \langle \phi_{\alpha,s\mathbf{k}_{uc}^1}^{val} \phi_{\beta,t\mathbf{k}_{uc}^2}^{cond} | K | \phi_{\gamma,y\mathbf{k}_{uc}^3}^{val} \phi_{\lambda,z\mathbf{k}_{uc}^4}^{cond} \rangle, \qquad (S4)$$

where $\alpha, \beta, \gamma$ and $\lambda$ are layer indices, $i$ and $p$ are pristine valence band indices, $j$ and $q$ are pristine conduction band indices. The pristine states $i\mathbf{k}_m, j\mathbf{k}_m, p\mathbf{k}'_m$ and $q\mathbf{k}'_m$ in the moiré BZ are related to $s\mathbf{k}_{uc}^1, t\mathbf{k}_{uc}^2, y\mathbf{k}_{uc}^3$ and $z\mathbf{k}_{uc}^4$ in the unit-cell BZ by band folding. The kernel matrix element is in Eq. S3 refers to the pristine supercell matrix element, while the matrix element in Eq. S4 refers to the pristine unit-cell matrix element. The screened Coulomb interaction used in Eq. S4 is that of pristine AB or 2H stacking. Thus, we can express each moiré kernel matrix element as a linear combination of pristine unit-cell kernel matrix elements.

The BSE kernel[23] is given by $K = -K_d + K_x$, where $K_d$ is a direct matrix element responsible for the attractive electron-hole interaction, and $K_x$ is a repulsive exchange interaction. The kernel matrix elements are six-dimensional integrals,

$$\left\langle \Phi_{\alpha,i\mathbf{k}_m}^{val} \Phi_{\beta,j\mathbf{k}_m}^{cond} | K^d | \Phi_{\gamma,p\mathbf{k}'_m}^{val} \Phi_{\lambda,q\mathbf{k}'_m}^{cond} \right\rangle$$

$$= \int d\mathbf{r} d\mathbf{r}' \Phi_{\beta,j\mathbf{k}_m}^{cond*}(\mathbf{r}) \Phi_{\lambda,q\mathbf{k}'_m}^{cond}(\mathbf{r}) W^{AB}(\mathbf{r},\mathbf{r}') \Phi_{\alpha,i\mathbf{k}_m}^{val}(\mathbf{r}') \Phi_{\gamma,p\mathbf{k}'_m}^{val*}(\mathbf{r}'), \qquad (S5)$$

$$\left\langle \Phi_{\alpha,i\mathbf{k}_m}^{val} \Phi_{\beta,j\mathbf{k}_m}^{cond} | K^x | \Phi_{\gamma,p\mathbf{k}'_m}^{val} \Phi_{\lambda,q\mathbf{k}'_m}^{cond} \right\rangle$$

$$= \int d\mathbf{r} d\mathbf{r}' \Phi_{\beta,j\mathbf{k}_m}^{cond*}(\mathbf{r}) \Phi_{\alpha,i\mathbf{k}_m}^{val}(\mathbf{r}) W^{AB}(\mathbf{r},\mathbf{r}') \Phi_{\lambda,q\mathbf{k}'_m}^{cond}(\mathbf{r}') \Phi_{\gamma,p\mathbf{k}'_m}^{val*}(\mathbf{r}'). \qquad (S6)$$

The pristine kernel matrix elements that we need to compute can be classified in terms of the layer indices, $\alpha, \beta, \gamma$ and $\lambda$.

When all the indices refer to layer 'a' or layer 'b', the kernel matrix element is an intralayer kernel matrix element of layer 'a' or layer 'b', respectively, with a pristine AB stacking screened

Coulomb interaction. For the intralayer kernel matrix elements of layer 'a', we choose the screened Coulomb interaction with the periodicity of the layer 'a', i.e., $W_\mathbf{q}^{AB}(\mathbf{G}_a, \mathbf{G}'_a)$ where $\mathbf{G}_a$ and $\mathbf{G}'_a$ are reciprocal lattice vectors of layer 'a'. Similarly, for the intralayer kernel matrix elements of layer 'b' we choose the screened Coulomb interaction with layer 'b' periodicity: $W_\mathbf{q}^{AB}(\mathbf{G}_b, \mathbf{G}'_b)$. The choice of $W$ has important implications for the Umklapp processes involved. For the intralayer kernel matrix elements of layer 'a', only terms with $\mathbf{k}_{uc}^2 - \mathbf{k}_{uc}^1 = \mathbf{k}_{uc}^4 - \mathbf{k}_{uc}^3 + \mathbf{G}_a$ are non-zero,

$$\left\langle \Phi_{a,i\mathbf{k}_m}^{val} \Phi_{a,j\mathbf{k}_m}^{cond} | K | \Phi_{a,p\mathbf{k}'_m}^{val} \Phi_{a,q\mathbf{k}'_m}^{cond} \right\rangle$$
$$= \left\langle \phi_{\alpha,s\mathbf{k}_{uc}^1}^{val} \phi_{\beta,t\mathbf{k}_{uc}^2}^{cond} | K | \phi_{\gamma,y\mathbf{k}_{uc}^3}^{val} \phi_{\lambda,z\mathbf{k}_{uc}^4}^{cond} \right\rangle \delta_{\mathbf{k}_{uc}^2 - \mathbf{k}_{uc}^1, \mathbf{k}_{uc}^4 - \mathbf{k}_{uc}^3 + \mathbf{G}_a} \quad (S7)$$

The other combinations of $\alpha, \beta, \gamma$ and $\lambda$ refer to interlayer kernel matrix elements. Since layer 'a' and layer 'b' are separated by about 3 Å, and the wavefunctions of each layer exponentially decay in vacuum, we only compute the dominant interlayer interactions for the direct and exchange matrix elements. From Eq. S5, we find that the interlayer interaction is non-negligible for the direct kernel matrix element only when $\alpha = \gamma$ and $\beta = \lambda$. For example, the contribution of $\Phi_{\beta,j\mathbf{k}_m}^{cond*}(\mathbf{r})\Phi_{\lambda,q\mathbf{k}'_m}^{cond}(\mathbf{r})$ is small to the integral in Eq. S5 when the two conduction states are in different layers. Similarly, the interlayer interaction is non-negligible for the exchange kernel matrix element when $\alpha = \beta$ and $\gamma = \lambda$ (from Eq. S6). For the interlayer direct kernel interactions, we approximate the screened Coulomb interaction to have the periodicity of layer 'a' and we further include only long-range interactions in the in-plane direction ($G_x = G'_x = 0$ and $G_y = G'_y = 0$), while fully including local-field effects in the out-of-plane direction:

$$W_{\mathbf{q}}(\mathbf{G}, \mathbf{G}') \approx W_{\mathbf{q}}(\mathbf{G}, \mathbf{G}')\delta_{G_x, G'_x=0}\delta_{G_y, G'_y=0}, \tag{S7}$$

where **G** is a reciprocal lattice vector. This approximation is valid for the interlayer matrix elements since the hole and electron are in different layers, hence the in-plane local-field effect and short-wavelength modulations of W are not as important as the long-wavelength interactions.

## S7. Validation of the generalized pristine unit-cell matrix projection method

We validate the approximations described in the previous section for a 38.2° twisted bilayer $WS_2$ for which the BSE calculations can be explicitly performed. The 38.2° twisted bilayer $WS_2$ contains 42 atoms in the moiré unit-cell and the structure was relaxed as described above. The electronic band structure, including spin-orbit coupling, is shown in Fig. S9. The explicit BSE was computed with a $12 \times 12 \times 1$ k-point sampling of the moiré BZ and 2 valence and 2 conduction bands. The dielectric matrix was computed in the superlattice using ~600 empty states and a plane-wave energy cutoff of 3.5 Ry. As described above, we express two valence and two conduction band edge states in the moiré superlattice as a linear combination of 8 valence and 8 conduction states of each pristine monolayer ($n = 8$ in Eqns. S1 and S2). The projection of the constructed electronic states on the original DFT computed states, $\langle \psi_{v\mathbf{k}_m}^{cons} | \psi_{v\mathbf{k}_m} \rangle$, was found to be more than 90%.

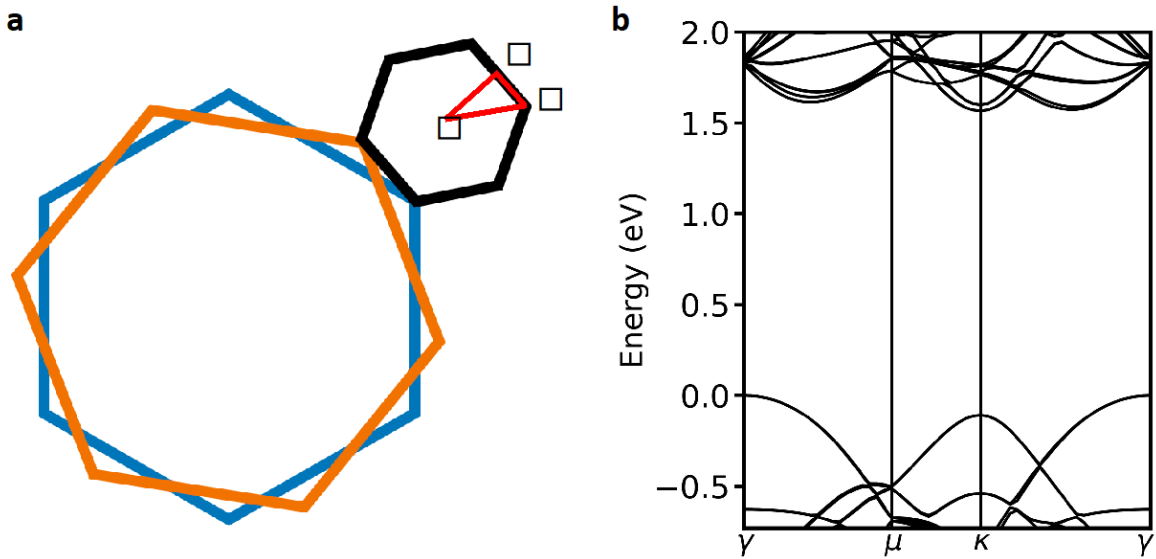

**Figure S9. Brillouin zones and band structure of 38.2° twisted bilayer moiré superlattice a.** Schematic of the Brillouin zones of pristine layer 'a' (blue), pristine layer 'b' (orange) and the 38.2° twisted bilayer moiré superlattice (black) **b.** Electronic band structure of the 38.2° twisted bilayer WS$_2$ plotted along the high-symmetry directions in the moiré BZ.

The $12 \times 12 \times 1$ k-point sampling of the superlattice BZ unfolds to a uniform $12\sqrt{7} \times 12\sqrt{7} \times 1$ k-point sampling in the pristine unit-cell BZ of each layer. The screened Coulomb interaction was computed in the pristine unit cell using this q-grid. We computed the electron-hole interaction kernel matrix elements using the generalized PUMP approach (Eq. S4) including all the approximations detailed in the above section. The absorption calculation was compared between the explicit and constructed BSE. The constructed absorption spectrum is in excellent agreement with the explicit calculation (Fig. S10), showing that our approximations are valid for a tolerance of about 10 meV in the excitation energies.

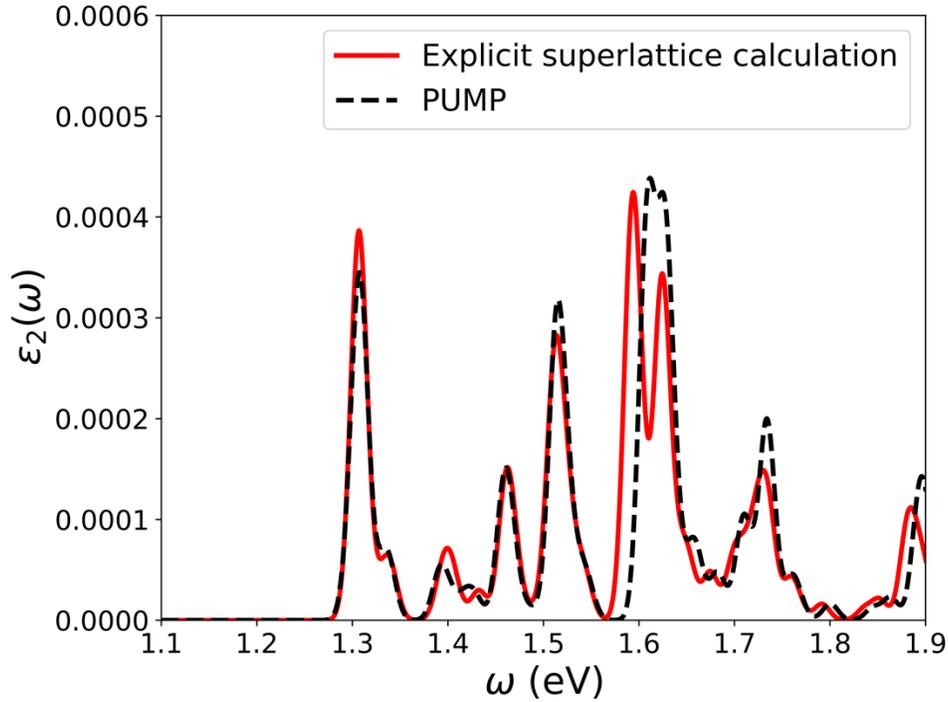

**Figure S10. Validation of the generalized PUMP approach.** Absorption spectrum of 38.2° twisted bilayer WS$_2$ computed by an explicit superlattice BSE calculation compared with the spectrum obtained using the pristine unit-cell matrix projection method.